\documentclass{osa-article}

\usepackage{siunitx}
\usepackage{physics}
\usepackage{hyperref}
\usepackage{mhchem}

\journal{oe}

\begin{document}
	
	\title{The optimal lattice depth on lifetime of D-band ultracold atoms in a triangular optical lattice}
	
	\author{Hongmian Shui,\authormark{1,2} Chi-Kin Lai,\authormark{1} Zhongcheng Yu,\authormark{1} Jinyuan Tian,\authormark{1}Chengyang Wu,\authormark{1} Xuzong Chen,\authormark{1} Xiaoji Zhou\authormark{1,2,3,*}}
	
	\address{\authormark{1}State Key Laboratory of Advanced Optical Communication System and Network, School of Electronics,
		Peking University, Beijing 100871, China\\
		\authormark{2}Institute of carbon-based thin film electronics, Peking University, Shanxi, Taiyuan 030012, China\\
		\authormark{3}Institute of Advanced Functional Materials and Devices, Shanxi University, Taiyuan 030031, China
	}
	
	\email{\authormark{*}xjzhou@pku.edu.cn}
	
	\begin{abstract}
		
		Ultracold atoms in optical lattices are a flexible and effective platform for quantum precision measurement, and the lifetime of high-band atoms is an essential parameter for the performance of quantum sensors. 
		In this work, we investigate the relationship between the lattice depth and the lifetime of D-band atoms in a triangular optical lattice and show that there is an optimal lattice depth for the maximum lifetime.
		After loading the Bose-Einstein condensate into D band of optical lattice by shortcut method, we observe the atomic distribution in quasi-momentum space for the different evolution time, and measure the atomic lifetime at D band with different lattice depths.
		The lifetime is maximized at an optimal lattice depth, where the overlaps between the wave function of D band and other bands (mainly S band) are minimized.
		Additionally, we discuss the influence of atomic temperature on lifetime.
		These experimental results are in agreement with our numerical simulations.
		This work paves the way to improve coherence properties of optical lattices, and contributes to the implications for the development of quantum precision measurement, quantum communication, and quantum computing.
	\end{abstract}
	
	\section{Introduction}
	With its high controllability, ultracold atoms in optical lattices provide a reliable platform for quantum precision measurement \cite{PhysRevLett.95.093202} and quantum simulation \cite{RevModPhys.78.179, RevModPhys.91.015005}, such as the atomic interferometer \cite{PhysRevLett.101.230801,PhysRevA.88.031605}, the quantum gate \cite{PhysRevA.104.L060601}, the observation of the quantum Hall effect \cite{Lohse2018,Tarnowski2019}, and the dynamics behavior of atoms in optical lattices \cite{PhysRevA.107.023303,PhysRevA.91.023623}.
	Nowadays, due to its different symmetry of wave function and novel interaction effect, the research on high-orbital atoms have attracted much attention \cite{Li_2016,Lewenstein2011}.
	Excited-band atomic interferometry is achieved at P and D band \cite{PhysRevA.101.023614}, the re-coagulation of P-band bosons is observed in honeycomb optical lattice with different symmetry \cite{PhysRevLett.126.035301,Wang2021}, and the quantum stripe ordering in triangular optical lattice is achieved \cite{PhysRevLett.97.190406}.
	
	To manipulate high-orbital atoms, it is crucial to increase their lifetime. 
	Some studies on lifetime of atoms in optical lattices have been carried out, including the development of echo technique to increase the lifetime of D-band atomic Ramsey interferometry \cite{Hu2018,Dong22}, and the measurement of decay mechanism \cite{Zhai2013,PhysRevA.90.013602,PhysRevLett.111.205302}.
	Lattice depth is one of the most important parameters to affect the atomic lifetime in optical lattices. For example, lattice depth influences the coherence of atoms in optical lattices, and brings about a quantum phase transition from a superfluid to a Mott insulator of ultracold atoms \cite{RN76}.
	Further, the eigen wavefunction of ultracold atoms in optical lattices changes with lattice depth \cite{Wirth2011}, and influences the two-body scattering channel \cite{PhysRevA.88.013608,PhysRevA.104.033326}, where two-body collision is the main decay mechanism of high-orbital atoms in optical lattices \cite{PhysRevA.72.053604,PhysRevA.88.013608}.
	However, the research on the relationship between lattice depth and the atomic lifetime at high band in two-dimensional (2D) optical lattices is rare.
	
	In this work, we study the influence of the lattice depth on the lifetime of D-band atoms in a 2D triangular optical lattice, and the atomic lifetime appears a maximum at the optimal lattice depth. We begin by loading a Bose-Einstein condensate (BEC) into the D band of a triangular optical lattice using our shortcut method. 
	We demonstrate the lifetime of D-band atoms at different lattice depths, and the lifetime appears a maximum at a lattice depth of around 4 Er, which is the result of the unique transformation of the D-band wave function. Meanwhile, the trend of lifetime varying with lattice depth is predicted by the scattering theory, which agrees well with our experimental results. Moreover, we measure the lifetime at the D band with different atomic temperatures, and show the square of atomic lifetime is inversely proportional to temperature, where lattice depth determines the proportional parameter.
	
	This paper is organized as follows. In Sec. 2, we describe the theoretical model for the lifetime of D-band atoms. In Sec. 3, our experimental procedure and the shortcut method for loading BEC into the D band of a triangular optical lattice are described. The experimental results of different lattice depths are described in Sec. 4. Finally, we give the discussion in Sec. 5 and the conclusion in Sec. 6.
	
	\section{Theoretical model for the lifetime of atoms in a triangular optical lattice}\label{sec:theoreticalModel}
\begin{figure}[htbp]
		\centering
		\includegraphics[width=0.9\linewidth]{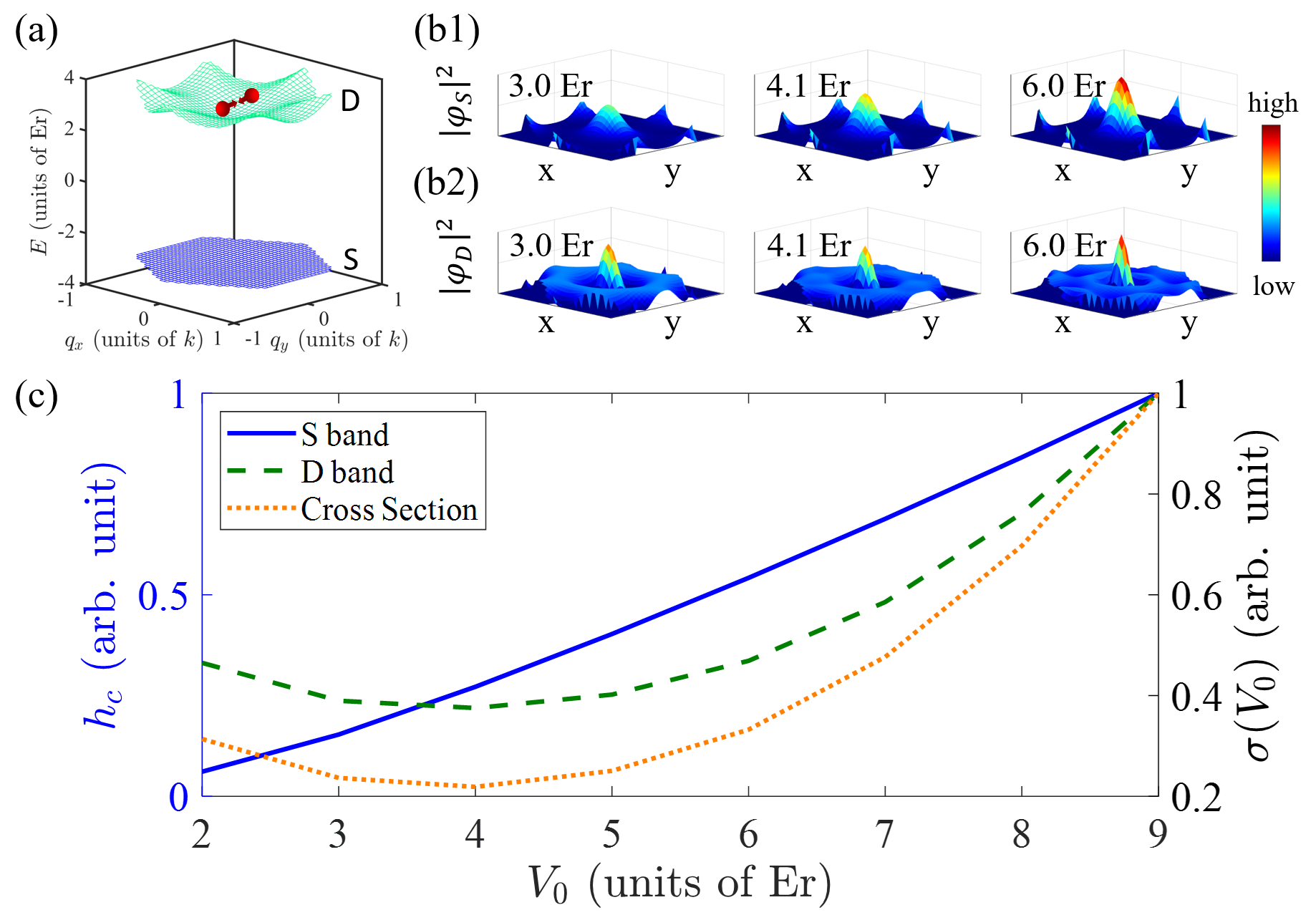}
		\caption{(a) Band structure of the triangular lattice with $V_0=5.0$ Er. The green and blue curved surfaces represent the S and D band, respectively. The red spheres represent the atoms prepared to the D band, and the arrows illustrate their collision. $k=2\pi/\lambda$, and $\lambda$ is the wavelength of the triangular optical lattice. (b1) and (b2) show the modulus squared of S- and D-band wave functions at different lattice depths.  (c) The blue solid and green dashed lines are the central peak height of normalized S- and D-band density under different lattice depths, respectively. The orange dotted line is the normalized total scattering cross section.
		}
		\label{fig:schematic}
	\end{figure}

	The lifetime of high-band atoms is an important parameter to improve the performance of quantum precision measurement and quantum simulation in optical lattices.
	In this work, we focus on studying the influence of lattice depth on the lifetime of the D-band atoms in a triangular optical lattice, as shown in Fig.\ref{fig:schematic}.(a).
	Collision is the essential mechanism responsible for the decay of ultracold atoms at high bands in optical lattices.
	For ultracold Bosons, the s-wave collision approximation is valid, as higher-order collisions (e.g., d-wave collision) can be neglected at the ultracold atoms regime. 
	In addition, three-body collisions are suppressed in the condensed phase \cite{Burt1997}. 
	Furthermore, collisions between ultracold atoms and thermal atoms can be neglected in high vacuum condition.
	As a result, the two-body s-wave collisions between ultracold atoms are the major factor.
	The Hamiltonian for the two-body scattering in a triangular optical lattice is given by
	\begin{align}
		\hat{H}&=-\frac{\hbar^{2}}{m}\frac{\partial^{2}}{\partial z^{2}}+\sum_{i=1,2}\left[-\frac{\hbar^{2}}{2m}\frac{\partial^{2}}{\partial x_{i}^{2}}-\frac{\hbar^{2}}{2m}\frac{\partial^{2}}{\partial y_{i}^{2}}+\sum_{j=1,2,3}\frac{1}{2}V_0\cos(\vec{k_j}\cdot\vec{r_i})\right]+U(\vec{r}),
		\label{Hamiltonian}
	\end{align}
	where  $\vec{r}_{1}$,\ $\vec{r}_{2}$ are the coordinates of two atoms, $\vec{r}=\vec{r}_{2}-\vec{r}_{1}$ is the relative coordinate of two atoms, $\vec{k}_{1,2,3}$ are the wave vectors of the triangular optical lattice, and $V_0$ is lattice depth. The two-body interaction $U(\vec{r})=\dfrac{4\pi \hbar^{2} a_s}{m}\delta(\vec{r})\dfrac{\partial}{\partial r}(r\cdot)$ is the Huang-Yang pseudo-potential characterized by the s-wave scattering length $a_{s}$, where $\delta(\vec{r})$ is the delta function. 
	
	The two-body scattering rate $R$ is dependent on the total scattering cross section $\sigma(V_0)$ and the overall velocity of atoms $v$ \cite{Zhai2013}:
	\begin{align}
		R=\sigma(V_0)v.
		\label{rate}
	\end{align}
	Further, the total scattering cross section is composed by scattering cross sections of different scattering channels $\sigma(V_0)=\sum_{(n'_1,n'_2)\neq (d,d)}\sigma (n'_1,n'_2)$, where $\sigma (n'_1,n'_2)$ is the cross section of scattering channel from $\Gamma$ point (zero quasi-momentum) at D band to quasi-momentum $\pm \vec{q}$ at $n'_1,\ n'_2$ band. 
	Following the derivation from \cite{PhysRevA.104.033326, Zhai2013}, the two-body scattering cross section of a channel is proportional to the overlapping integral
	\begin{align}
		\sigma(n'_1,n'_2) \propto \int \mathrm{d}\vec{q}\, \left|\int \text{d}\vec{r} u_{n'_{1},\vec{q}}^{*}(\vec{r})u_{n'_{2},-\vec{q}}^{*}(\vec{r})u_{n_{1},0}(\vec{r})u_{n_{2},0}(\vec{r})\right|^{2},
		\label{overlap}
	\end{align}
	where $u_{n_{i},\vec{q}_{i}}(\vec{r}_i)\ (i=1,2)$ is the single-atom eigenstate of Hamiltonian Eq.(\ref{Hamiltonian}) in band $n_{i}$ at quasi-momentum $\vec{q}_{i}$. And $u_{n_{i},\vec{q}_{i}}(\vec{r}_i),\ u_{n'_{i},\vec{q}^{\, '}_{i}}(\vec{r}_i)$ denote the eigenstates of initial and final states of corresponding atoms.
	By defining the lifetime $\tau_D$ of the D-band atoms as the point at which the atomic number density drops to $\exp(-1)$, the atomic lifetime at D band is given by \cite{PhysRevA.104.033326,Zhai2013}
	\begin{align} 
		\label{eq:tau vs T}
		\tau_D = \frac{\exp(1)-1}{n_0\sigma(V_0)v},
	\end{align}
	where $n_0$ is defined as the atomic number density of D-band atoms at $t=0$.
	
	The overlapping integral is determined by both initial and final states, which are dependent on the lattice depth $V_{0}$.
	To analyze the influence of lattice depth on the atomic lifetime at D-band, we simplify the problem by only considering the dominant scattering channel $(d,d)\rightarrow(s,s)$ in the triangular optical lattice \cite{PhysRevA.104.033326}, and research the overlap of wave function at S and D band with different lattice depths.
	Fig.\ref{fig:schematic}.(b1) and (b2) show the density (modulus squared of wave functions) in real space for both the S band and D band at the $\Gamma$ point with $V_0= 3.0,\ 4.1,\ 6.0$ Er. 
	For $V_0=3.0 \ \rm Er$, the S-band density displays a Gaussian-like shape, and spreads out in real space. In contrast, the D-band density concentrates at the center. For $V_0=4.1 \ \rm Er$, the S-band density becomes narrowing due to the transformation from a plane wave form to a localized form of S-band wave function, yielding a narrower Gaussian-like shape, while the central peak of D-band density is minimized and the density spreads out to the maximum extend, resulting in minimal overlap of the two wave functions at that point. For $V_0=6.0 \ \rm Er$, the S-band density exhibits an even narrower Gaussian-like shape, and the D-band density redistributes at the center.  
	We study the overlap between the wave functions of S band and D band by extracting the height of the central peak $h_c$ of the distribution density. 
	The height of the central peak of the D-band density performs a trend that first decreases and then increases, and is minimized at $V_0=4.1$ Er, while the height of central peak of the S-band density increases monotonically with the increasing lattice depth, as the blue solid and green dashed lines shown in Fig.\ref{fig:schematic}.(c). 
	
	The non-monotonic change of D-band wave function results in the optimal lattice depth for minimum scattering cross section and maximum lifetime.
	The total scattering cross section of the initial state with $\Gamma$ point at D band in the triangular lattice with lattice depth increasing is calculated by the scattering theory, as the orange dotted line shown in Fig.\ref{fig:schematic}.(c). 
	The total scattering cross section has a similar trend that first decreases and then increases, and the lifetime of D-band atoms $\tau_D$ has a maximum with an optimal lattice depth $V_0=4.1$ Er.
		
	Another factor that influences the scattering rate is the velocity of atoms $v$, which can be decomposed into the thermal velocity $v_{\rm t}$ and the group velocity $v_{\rm g}$, as $v=\sqrt{v^{2}_{\rm t}+v^{2}_{\rm g}}$.
	The thermal velocity follows a Maxwell-Boltzmann distribution, $v_{\rm t}=\sqrt{3k_{\rm B}T/m}$, where $k_{\rm B}$ is the Boltzmann constant and $T$ is the temperature of the atoms. As the temperature of atoms increases, the thermal velocity of atoms also increases, and atoms are expected to have a greater scattering rate.
	The group velocity is defined as $v_{\rm g} = \partial_{k_{\perp}} E/\hbar$, where $E$ is the energy of atoms and $k_{\perp}$ is the wave vector perpendicular to the cross section. 
	Since the atoms are prepared in the vicinity of D-band $\Gamma$ point, the group velocity $v_{\rm g}$ is much smaller than $\hbar k/m$, where $k=2\pi/\lambda$ and $\lambda$ is the wavelength of the optical lattice. Given that the temperature of BEC is approximately $100\ \rm nK$, $v_{\rm t} \approx \hbar k/m$, which indicates the group velocity $v_{\rm g}\ll \hbar k/m\approx v_{\rm t}$, the group velocity can be neglected. From the Eq.(\ref{eq:tau vs T}), $\tau_D^2$ is inversely proportional to temperature $T$.
	
	\section{Experiment description}\label{sec:experimentImplementation}
		\begin{figure}
		\centering
		\includegraphics[width=0.85\linewidth]{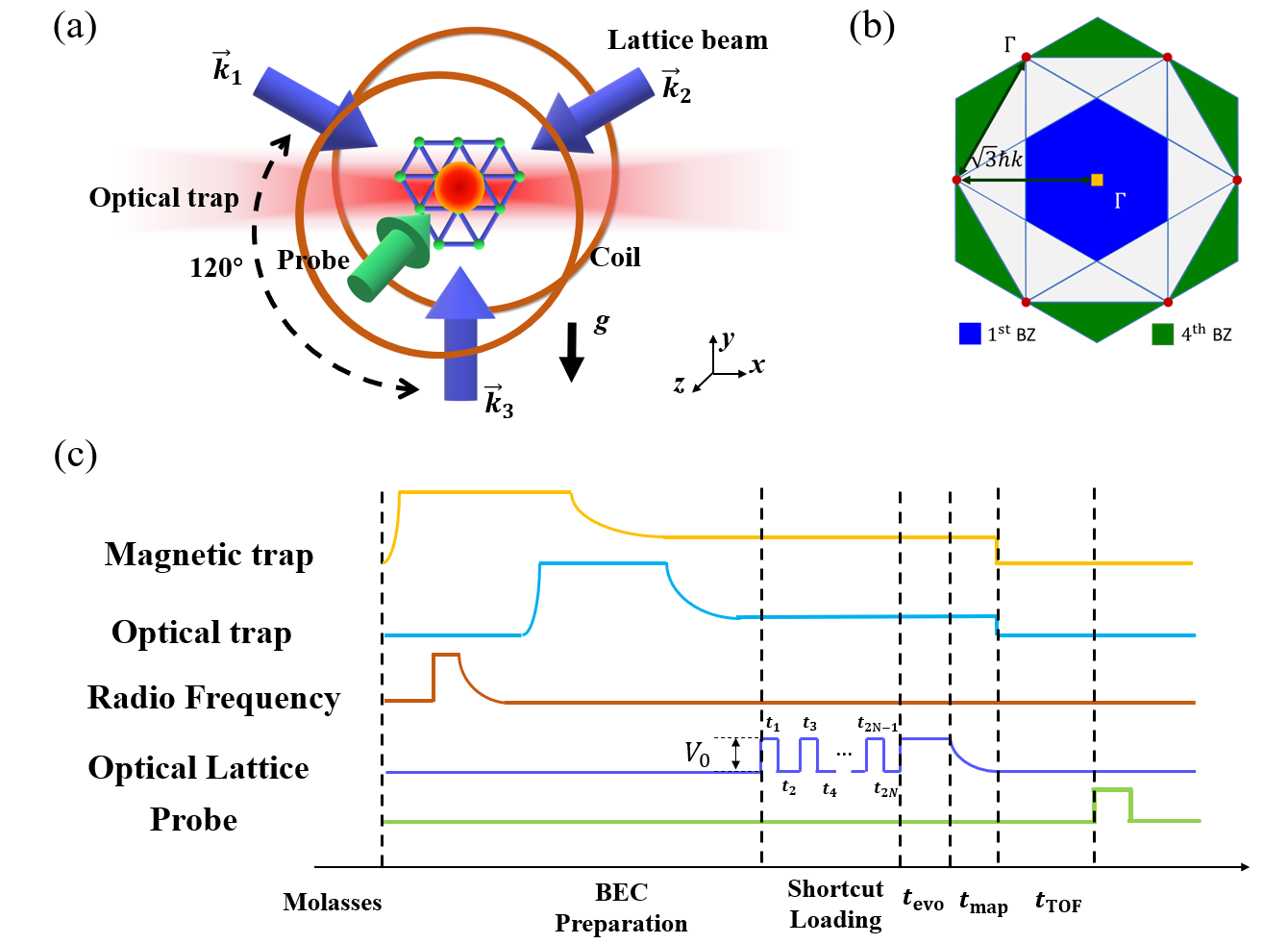}
		\caption{(a) The schematic of the experimental setup. The brown circles in the $x$-$y$ plane represent the magnetic trap coils and the red Gaussian beam along the $x$-axis marks the optical trap. The three blue arrows in the $x$-$y$ plane stand for the lattice beams with an enclosing angle of $120^\circ$, of which $k_j(j = 1,2,3)$ represent wave vectors. The green arrow is the probe beam, which is perpendicular to the lattice plane. (b) Schematic diagram of the Brillouin zone (BZ) of the triangular optical lattice. Blue area for the first BZ and green area for the fourth BZ. $\Gamma$ points of the first and the fourth BZ are marked as the yellow square and the red points, respectively. (c) A typical experimental sequence. The yellow line and light blue line represent the magnetic trap and optical trap. The brown line represents the radio frequency (RF) cooling. The blue line represents the optical lattice, and several pulses before $t_{\rm evo}$ are shortcut pulses designed to load BEC into the D band with lattice depth $V_0$. After an evolution time $t_{\rm evo}$, the lattice beam intensity decreases to zero adiabatically in time $t_{\rm map}$. The green line represents the probe beam, which is turned on for absorption imaging after the time-of-flight process with time $t_{\rm TOF}$. The horizontal axis indicates the experimental stages.}
		\label{fig:experimentalSetup}
	\end{figure}
	Fig.\ref{fig:experimentalSetup}. (a) shows the schematic of the experimental set-up, and the experimental time sequence is presented in Fig.\ref{fig:experimentalSetup}. (c). To prepare the BEC, we use a magnetic trap to confine the molasses and then cool the atoms via radio frequency (RF) cooling. Then, the magnetic trap falls adiabatically and the atoms are transformed into an optical trap. Finally, through the evaporation of atoms in the optical trap, we obtain the BEC with about $2\times 10^5$ atoms in the $|F = 2,m_F  = +2\rangle $ state. By controlling the final intensity of the optical trap at the evaporation process, we can change the temperature of BEC from 50 nK to 200 nK, while we modulate the density of molasses to keep the atomic number of BEC unchanged. After the preparation, the BEC is confined in a hybrid trap consisted by the magnetic trap and the optical trap with the harmonic trapping frequencies $(\omega_x,\omega_y,\omega_z ) = 2\pi \times (28,55,60)$ Hz.
	
	Then we load the BEC into the D band of the triangular optical lattice by the non-adiabatic shortcut loading method.
	The 2D triangular optical lattice with tube-shaped lattice sites is formed by three intersecting laser beams with $120^\circ $ enclosing angles, and their polarizations are perpendicular to the lattice plane ($x$-$y$ plane) constructed by three wave vectors $\vec{k}_j\ (j = 1,2,3)$, as shown in Fig.\ref{fig:experimentalSetup}.(a), where $\left|\vec{k}_j\right| =2\pi /\lambda$ and $\lambda=1064$ nm is the wavelength of the triangular optical lattice. After being loaded into the D band, the BEC in the optical lattice evolves for a period of time $t_{\rm evo}$. Then, the lattice beams are adiabatically turned off in the form $\exp(-t_{\rm map}/\tau)$ with the mapping time $t_{\rm map}  = 1\ \mathrm{ms}$ and time constant $\tau = 200\ \mathrm{\rm\mu s}$. In this process, the atoms populated in $\mathrm{n^{th}}$ band are mapped to the $\mathrm{n^{th}}$ BZ. Finally, the absorption images are taken after the time-of-flight (TOF) process with $t_{\rm TOF}  = 32\ \mathrm{ms}$ to measure the distribution of atoms in each band.
	
	In the experiment, we apply the non-adiabatic shortcut method to load the BEC particularly at the $\Gamma$ point of the D band in the triangular optical lattice, as shown in Fig.\ref{fig:experimentalSetup}.(b). 
	The shortcut method modulates the atomic states by turning on and off the lattice beams in a designed time sequence. The atomic state evolves with different Hamiltonians when the lattice beams are switched on and off. By well designing the duration and interval time sequences of the lattice pulses, the shortcut method is optimized to reach the target state with high fidelity \cite{Zhou_2018}. For the lattice depth $V_0$ of 5.0 Er, the on/off time of the four-pulse sequence is $29.5/15.5/16.5/29.5/6.5/31.0/18.0/12.5\ \mathrm{\mu s}$, as shown in Fig.\ref{fig:experimentalSetup}.(c).
	The theoretical fidelity $F^T$ of the sequence reaches $99.60\%$, and the experimental fidelity $F^E$ reached $98.1\%$. Table 1 shows the shortcut sequences used in the experiments.
	
	\begin{center}
		Table 1. The shortcut sequences of loading atoms into the D band from BEC with different $V_0$.
		\setlength\tabcolsep{3pt}
		\begin{tabular}{cllllllllll}
			\hline
			$V_0$ (Er)& $t_1$ ($\mathrm{\mu s}$)&$t_2$&$t_3$&$t_4$&$t_5$&$t_6$&$t_7$&$t_8$&$F^T$&$F^E$\\ 
			\hline
			3 &13.50&11.50&49.00&9.50&8.50&56.50&11.00&11.00&99.95\%&94.6\%\\
			4 &27.25&18.29&12.51&30.04&0.17&49.97&14.01&12.93&99.70\%&95.9\%\\
			5 &29.50&15.50&16.50&29.50&6.50&31.00&18.00&12.50&99.60\%&98.1\%\\
			6 &24.50&17.50&15.50&30.50&6.50&31.00&18.00&12.50&99.20\%&97.2\%\\
			8 &10.50&24.50&18.50&13.00&10.50&55.50&12.50&10.00&99.30\%&96.8\%\\
			\hline
		\end{tabular}
		\label{Table shortcut}
	\end{center}
	
	To modulate the lattice depth and improve the system's stability, two acousto-optic modulators (AOM) are used to control the triangular optical lattice. The first AOM modulates the lattice depth and is also used for laser intensity feedback control, achieving stability of $0.2\%$ @ 1s. 
	The second AOM, which is manipulated by an RF switch, works for producing a fast switch of optical lattice depth that is utilized in implementing the D-band loading shortcut sequence.
	The relative phase drift of three laser beams is mainly caused by the frequency drift of the laser. Due to the short lattice working time ($<100$ ms in a single experiment), the phase drift is about $1\times 10^{-4}$ $\pi$ , which is a negligible value. The pressure inside the vacuum chamber during the experiment is approximately $6\times 10^{-9}\ \mathrm{Pa}$, excluding the condition of collision between ultracold atoms and background. For example, at this vacuum degree, the lifetime of S band atoms can reach 2 s, which is much longer than the lattice working time.
	
	\begin{figure}[htbp]
		\centering
		\includegraphics[width=0.9\linewidth]{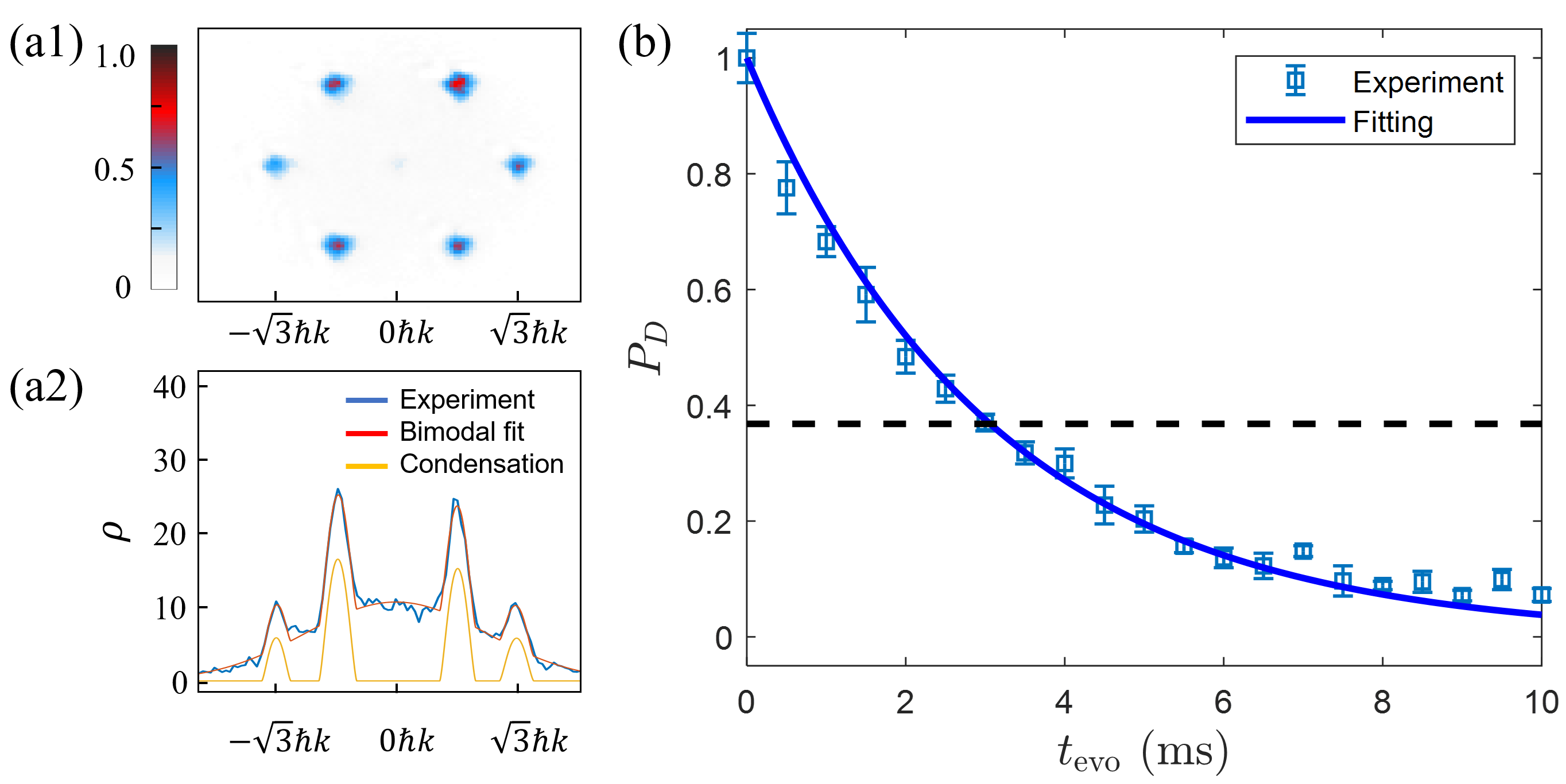}
		\caption{(a1) Typical TOF pattern of the experiment. (a2) Fitting method of this experimental image. By summing the 2D image along the y direction, we obtain the density distribution of atoms. The bimodal fit is performed to extract the number of D-band atoms. Codes for data analyzing are shown in Code 1 (Ref.\cite{code_ana}). (b) Measurement of atomic lifetime at the D band. The points of square represent experimental results, and the blue solid curve is the fitting result. The intersection of the black dashed line and the blue solid curve indicates the D-band lifetime.}
		\label{fig:ExperimentalResult}
	\end{figure}
	
	\section{Experimental results}\label{sec:experimentResult}
	\subsection{Measurement of Atomic lifetime at D band}
	In the experiment, the distribution of atoms in quasi-momentum space is observed after band mapping, where the atomic distribution in  $\mathrm{n^{th}}$ band is mapped to the area of  $\mathrm{n^{th}}$ BZ. Fig.\ref{fig:ExperimentalResult}.(a1) shows a typical band mapping image after shortcut loading atoms into D band. The atoms split into six clusters and condensate at the $\Gamma$ point of $\mathrm{4^{th}}$ BZ. Here, we use a bimodal fitting method to obtain the number of atoms in the D band, as shown in Fig.\ref{fig:ExperimentalResult}.(a2). 
	The atomic density in the TOF image is summed along the y-axis, compressing the 2D image to 1D function. After summation, the six clusters of atoms become four clusters along the $x$-axis. Then, we extract the atomic density in the D band by the bimodal fitting function $\rho$, which is given by
	\begin{align}
		\rho=\sum_{i=1}^{4}a_i\left(1-\frac{(r-r_i)^2}{w_i^2}\right)^{3/2}+b \exp\left(-\frac{(r-r_0)^{2}}{\sigma^{2}}\right).
		\label{bimodel}
	\end{align}
	The first term fits the atoms condensing at D band with amplitude $a_i$ and width $w_i$, while the second term fits thermal atoms with amplitude $b$ and width $\sigma$. The centers of the four atomic clusters and thermal atoms are given by $r_i$ and $r_0$. In Fig.\ref{fig:ExperimentalResult}.(a2), the blue, red, and yellow solid lines represent experimental data, bimodal fitting result, and fitted condensation, respectively. By summing and normalizing the fitted condensation result, we get the proportion of atoms at D band $P_D$.
	
	To measure the atomic lifetime at the D band, we extract the number of D-band atoms with different evolution time $t_{\rm evo}$. Fig.\ref{fig:ExperimentalResult}.(b) shows the typical experimental result, where the temperature of BEC is 150$\ \mathrm{nK}$ and the lattice depth is 5 Er. The blue squares are experimental results, and the error bar shows the standard error of five measurements. 
	With the evolution time increasing, the number of D-band atoms gradually decays, and we use the exponential curve to fit the atomic proportion-time curve. 
	The time for fitting curve decrease to the $\exp(-1)$ is taken as the lifetime, and the D-band lifetime $\tau_D$ is $3.06\pm 0.16$ ms at such condition.
	
	\subsection{Atomic lifetime at D band under different lattice depths}
	\begin{figure}[htbp]
		\centering
		\includegraphics[width=0.9\linewidth]{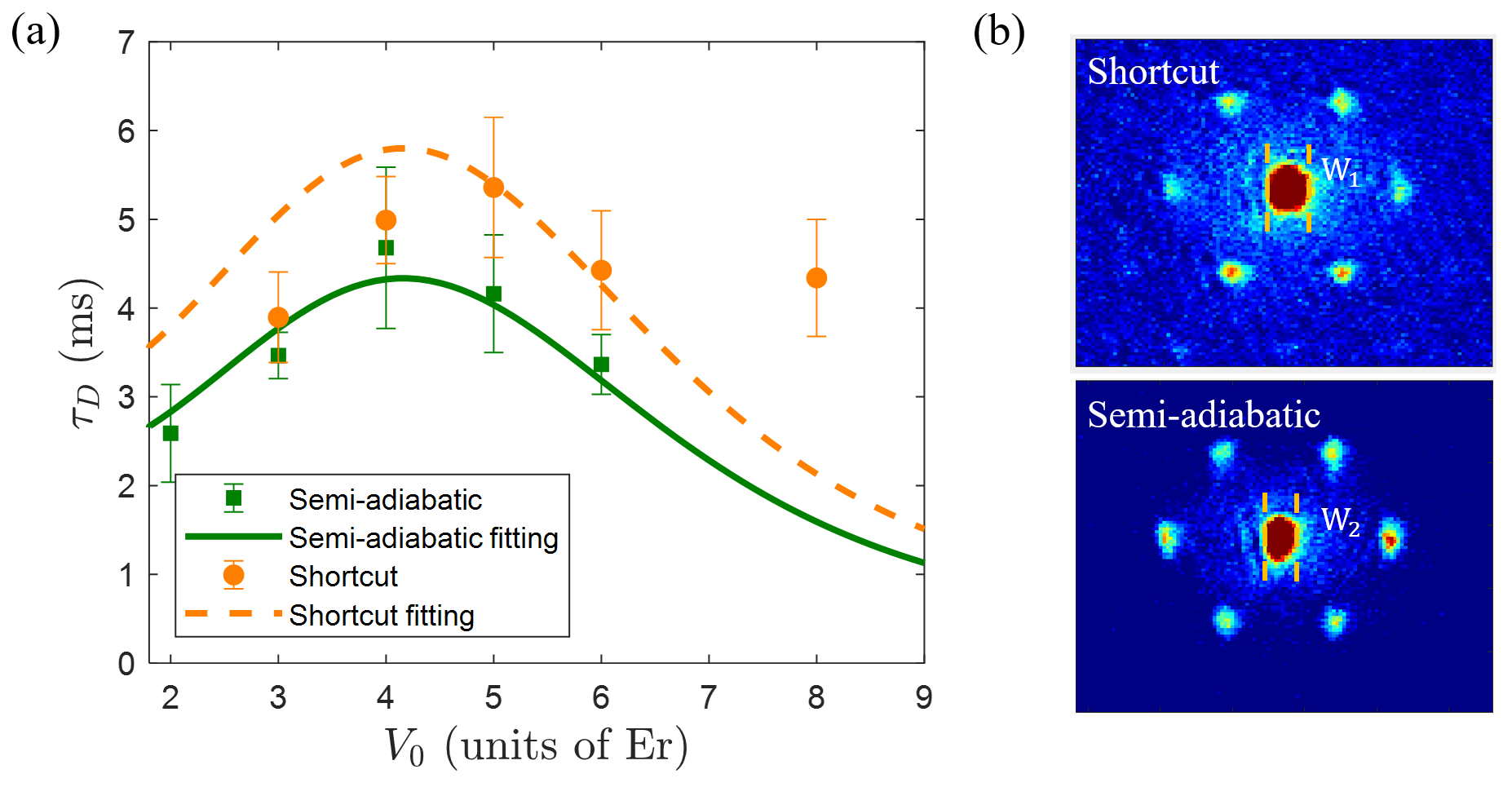}
		\caption{Experimental results under different lattice depths. (a) The points of the orange circle and green square represent shortcut and semi-adiabatic experimental results, respectively. The results for each point are averaged from five experiments. The orange dashed line and green solid line are the corresponding theoretical fitting results. (b) TOF patterns of atoms at the D band of triangular lattice with different loading methods. The upper and lower two patterns are the results of shortcut and semi-adiabatic loading, and $\rm W_1$ and $\rm W_2$ are the widths of their momentum distribution, respectively.}
		\label{fig:DifferentV}
	\end{figure}
	To study the change of the D-band lifetime $\tau_D$ under different lattice depths, we perform experiments with several lattice depths $V_0$ at fixed BEC temperature $T=100$ nK, where the condensation proportion of atoms is about 43\%. The experimental results obtained through the shortcut loading method are presented as the orange circles in Fig.\ref{fig:DifferentV}.(a), where the orange dashed line is the corresponding theoretical fitting result, and the root mean square error (RMSE) between experimental data and fitting is 1.72. With the lattice depth increasing, the lifetime first increases and then decreases, and the maximum lifetime occurs at around 5 Er, which has a deviation from the theoretical predicted lattice depth at 4.1 Er. 
	We attribute this discrepancy to the imperfect experimental fidelity. In the experiment, due to the imbalance of three lattice beams and the non-ideal square-wave shortcut pulses, the experimental fidelity can not reach the theoretical fidelity, as shown in Table 1. Fig.\ref{fig:DifferentV}.(b) shows the typical imaging of atoms after shortcut loading without band mapping. By fitting the imaging in Fig.\ref{fig:DifferentV}.(b), we find that the condensation proportion of D-band atoms after shortcut loading reduces to 27.6\%, while the width of momentum distribution is $\rm W_1=0.71\pm 0.04\ \hbar k$. The broadening of momentum distribution causes the D-band atoms to occupy different quasi-momentum states with $\vec{q}\neq0$.
	The scattering cross section changes with the initial state of D-band atoms $u_{d,\vec{q}}(\vec{r})$, resulting in a shift of optimal lattice depth corresponding to maximum experimental lifetime. 
	The interplay between the reduction of condensation proportion and the wider momentum distribution causes such deviation.
	
	To improve the D-band loading process, we use the semi-adiabatic loading method. This method is to adiabatically load the BEC into the S band of the triangular optical lattice at first, and then transfer the atoms into the D band by shortcut method. The average experimental fidelity of semi-adiabatic is 97.2\%, which is higher than that of shortcut loading with 96.5\%. 
	The higher experimental fidelity makes the atoms after the semi-adiabatic loading have greater condensation proportion and less broadening of momentum distribution.
	The typical imaging of semi-adiabatic loading at 5 Er, 100 nK without band mapping is shown in Fig.\ref{fig:DifferentV}.(b), where the condensation proportion is 34.5\% and the width of momentum distribution is $\rm W_2=0.53\pm 0.02 \ \hbar k$. After the semi-adiabatic loading process, we measure the atomic lifetime of the D band in different lattice depths, and the experimental results are represented by green squares in Fig.\ref{fig:DifferentV}.(a). The corresponding fitting result is represented by green solid line, and the RMSE between experimental data and fitting is 0.97, indicating a better agreement with the theoretical prediction.
	
	In summary, the enhanced fidelity of the D-band loading leads to an increment in the condensation proportion, while reducing the width of the momentum distribution, which contributes to the discrepancy between the experiment and theory. Specifically, the fidelity improvement results in a reduction of the momentum broadening by approximately 25\% after applying the semi-adiabatic shortcut method. Qualitatively speaking, this extent of reduction has sufficiently fixed the shift of the lattice depth corresponding to the maximum lifetime.

	\section{Discussion}\label{sec:Discussion}
	\subsection{Atomic lifetime at D band with different temperatures}
	\begin{figure}[htbp]
		\centering
		\includegraphics[width=0.9\linewidth]{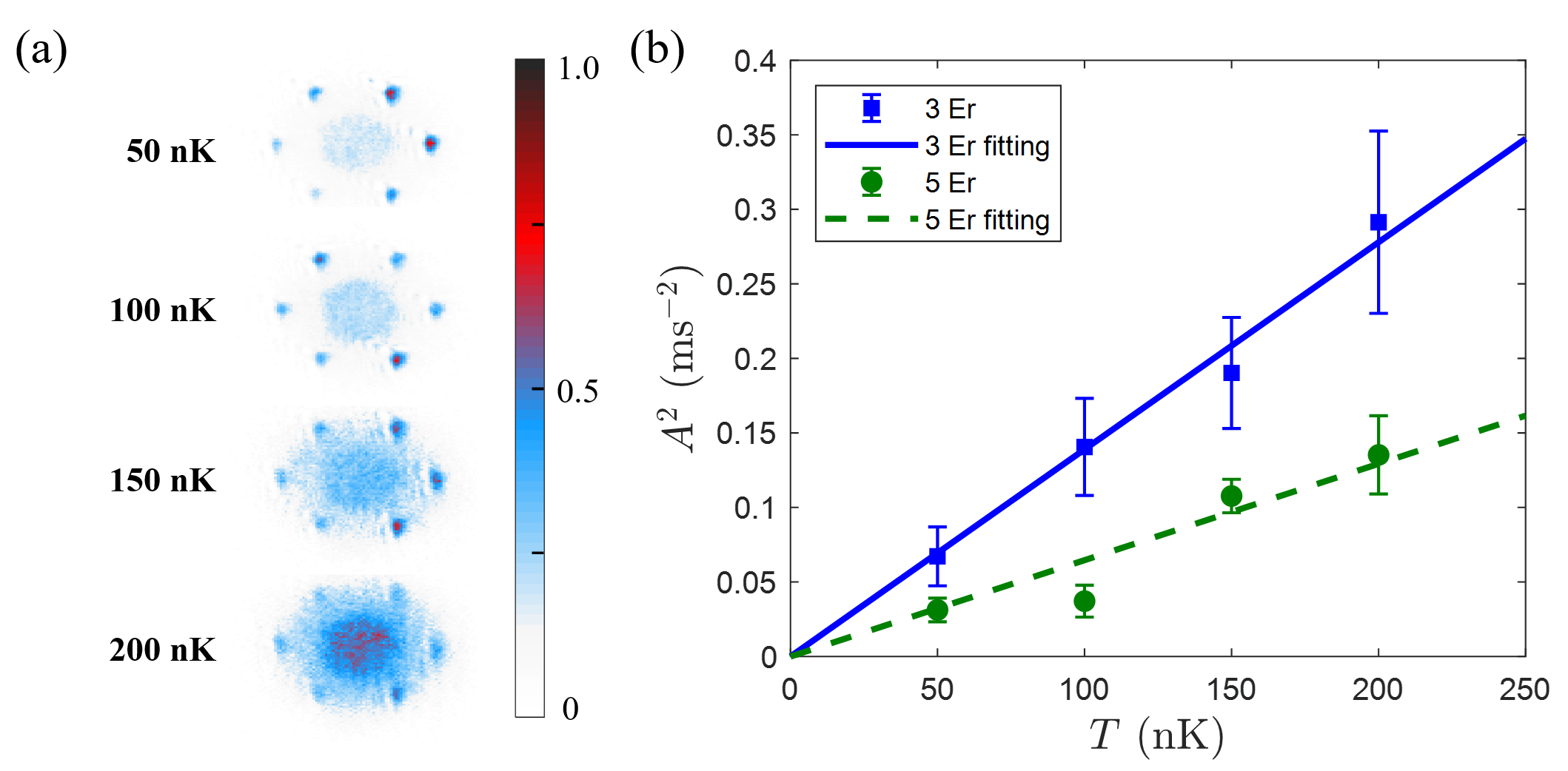}
		\caption{The atomic lifetime at D band of optical lattice for the different temperatures. (a) TOF patterns of different temperatures with $t_{\rm evo} = 2$ ms at $V_0=5$ Er obtained by shortcut loading method. (b) The measured lifetime of 3 Er (square points) and 5 Er (circle points) for the different temperatures, where each point is averaged by five times. The blue solid line and green dashed line represent the fitting results.}
		\label{fig:DifferentT}
	\end{figure}
	
	In order to further study the effect of lifetime in the D band of optical lattice, we investigate the influence of atomic temperature.
	We perform experiments at BEC temperatures $T=$ 50, 100, 150, and 200 nK with lattice depths $V_0=$ 3 and 5 Er. Fig.\ref{fig:DifferentT}.(a) shows the TOF images after 2 ms evolution at 5 Er. 
	With the temperature increasing, the decay of atoms condensing at D band becomes more apparent, and the number of scattered atoms increases.
	
	To quantitatively study the relationship between temperature and D-band lifetime, we define the decay rate of D-band atoms as $A=1/\tau_D$. Using the method shown in Sec 4.1, we measure the decay rate with different temperatures at 3 Er and 5 Er, and Fig.\ref{fig:DifferentT}.(b) shows the results.
	The data is fitted with the relation $A^2 = c T$ indicated by the theoretical model. The blue solid line and and the green dashed line represent the corresponding fitting results in Fig.\ref{fig:DifferentT}.(b). 
	The proportional parameter $c$ is proportional to the square of the scattering cross section, and is determined by lattice depth. Since the cross section of 5 Er is smaller than 3 Er, the proportional parameter $c$ of 5 Er is smaller than 3 Er.
	In Fig.\ref{fig:DifferentT}.(b), the proportional parameter $c$ of 5 Er is $64.6 \ \rm nK^{-1}ms^{-2}$, and $c$ of 3 Er is $138.9 \ \rm nK^{-1}ms^{-2}$.
	The experimental results show that the relationship between temperature and atomic lifetime at D band is explicitly described by $\tau_D\propto T^{-1/2}$, and the lattice depth determines the proportional parameter.
	
	\subsection{Maximum atomic lifetime at D band with different lattice configurations}
	\begin{figure}[htbp]
		\centering
		\includegraphics[width=0.9\linewidth]{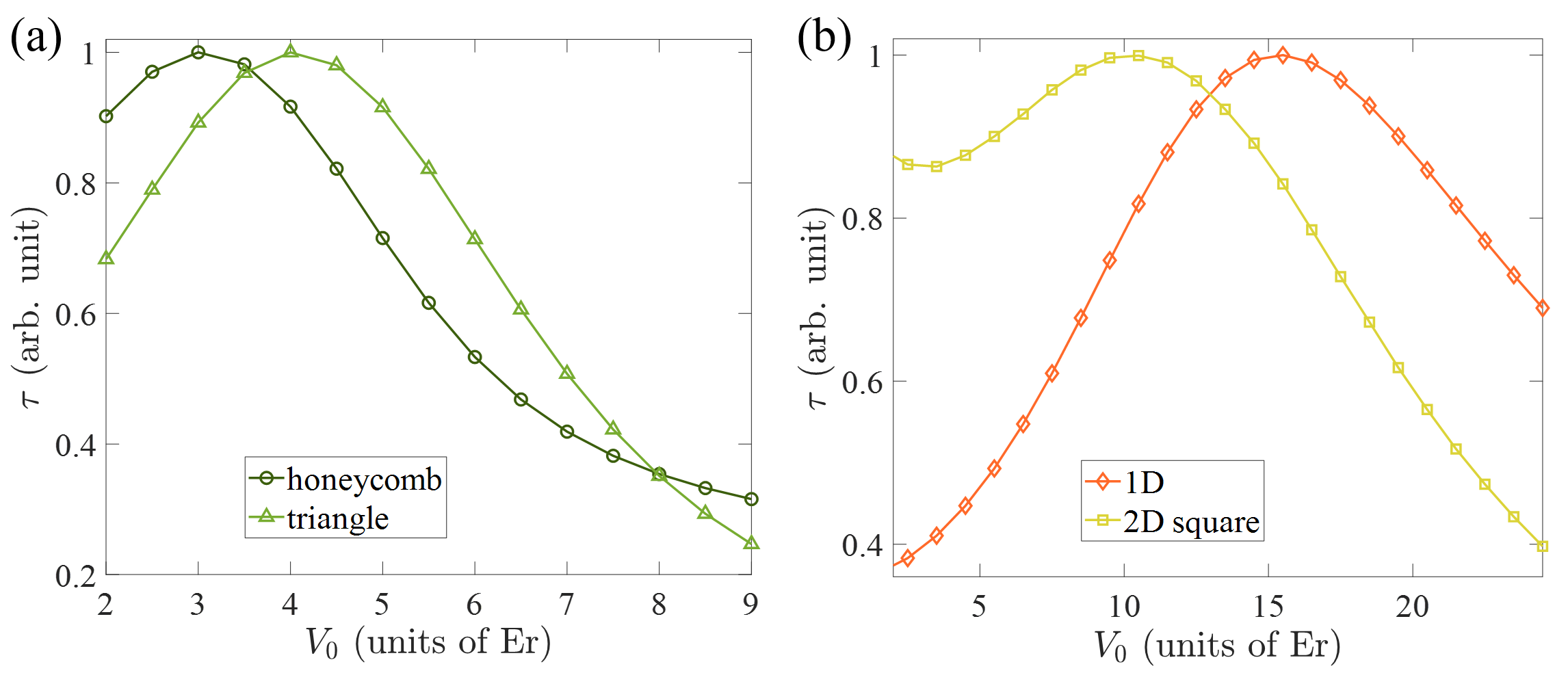}
		\caption{(a) The D-band lifetime of different lattices as a function of lattice depth, including honeycomb lattice (dark green circle) and triangular lattice (green triangle) in (a), and 1D lattice (orange diamond) and 2D square lattice (yellow square) in (b). These lines are normalized exclusively to the maximum lifetime of different lattices.}
		\label{fig:con}
	\end{figure}
	In the above experiment and theory, we have shown that the maximum lifetime of D-band atoms in triangular lattice is around $V_0=$ 4.1 Er. However, the maximum lifetime of D-band atoms by minimizing the overlap between the wavefunction of D band and other bands does not only occur in triangular lattice, but also exists in other structures of optical lattices. To demonstrate this phenomenon, we extend our numerical simulation to different lattice configurations, including 1D, 2D square, and honeycomb lattices. Codes for numerical simulation are shown in Code 2 (Ref.\cite{code_sim}).
	
	Similar to the case of triangular lattice shown in Fig.\ref{fig:schematic}.(b), the D-band densities in 1D, square and honeycomb lattice also exhibit first spreading and then concentrating feature with the lattice depth increasing, while the density of the lower bands monotonically change, which leads to a minimum of overlap and the maximum lifetime.
	Fig.\ref{fig:con}.(a) and (b) display the D-band lifetime as a function of lattice depth for these different lattice configurations, and the lattice depths for maximizing the lifetime of D-band atoms of these lattices are summarized in Table 2. In general, we conclude that the minimum of the overlapping integral between the D band and other bands exists due to the non-monotonic transformation of the D-band wavefunction, yielding the maximum D-band lifetime.
	
	\begin{center}
		Table 2. The lattice depths $V_0$ with maximum lifetime of different lattice configurations.
		\begin{tabular}{lllll}
			\hline
			Lattice& Triangle & Honeycomb & 2D Square & 1D\\ 
			\hline
			$V_0$ (Er)&4.1&3.1&9.9&15.4\\
			\hline
		\end{tabular}
		\label{Table 2}
	\end{center}
	
	
	\section{Conclusion}\label{sec:Conclusion}
	In this work, we research the relationship between lattice depth and atomic lifetime at D band in a triangular optical lattice. Due to the change of D-band wave function with different lattice depths, the atomic lifetime of D band has a maximum value at an optimal lattice depth. Also, we demonstrate that the square of atomic lifetime at D band is inversely proportional to temperature, where lattice depth determines the proportional parameter. The experimental results agree with theoretical predictions.
	This work provides insight into how the lattice depth and temperature interplay with the coherence properties of optical lattices, and paves a way to improve the coherence of the optical lattice systems.
	
	\section*{Funding}
	National Key Research and Development Program of China (Grants No. 2021YFA0718300 and No. 2021YFA1400900);
	National Natural Science Foundation of China (Grants No. 11934002 and No. 11920101004);
	Science and Technology Major Project of Shanxi (Grant No. 202101030201022);
	Space Application System of China Manned Space Program.
	
	\section*{Disclosures}
	The authors declare no conflicts of interest.
	
	\section*{Data Availability} Data underlying the results presented in this paper are available in Code file 1, Ref.\cite{code_ana}, and Code file 2, Ref.\cite{code_sim}.
	
	\bibliography{ref}
	
\end{document}